\documentclass[12pt]{article}
\usepackage{amsmath, euscript, amssymb, amsfonts, latexsym}

\begin{document}

\begin{center}
{\Large\bf NON-LOCALIZABILITY AND ASYMPTOTIC

\vspace{0.5cm}
COMMUTATIVITY}

\vspace{1cm}
V.~Ya.~Fainberg and M.~A.~Soloviev

\vspace{1cm}
Department of Theoretical Physics P.N.Lebedev Physical Institute

Leninsky prospect, 53, 117924, Moscow, Russia\footnote{E-mail address:
theordep@sci.fian.msk.su}
\end{center}

\vspace{2.5cm}

\centerline{\bf ABSTRACT}
\begin{quotation}

        The mathematical formalism commonly used in treating nonlocal highly
singular interactions is revised. The notion of support cone is introduced which replaces that of
support for nonlocalizable distributions. Such support cones are proven to exist
for distributions defined on the Gelfand-Shilov spaces $S^\beta$, where
$0<\beta <1$ . This result leads to a refinement of previous generalizations of
the local commutativity condition to nonlocal quantum fields. For string
propagators, a new derivation of a representation similar to that of
K\"{a}llen-Lehmann is proposed. It is applicable to any initial and final
string configurations and manifests exponential growth of spectral densities
intrinsic in nonlocalizable theories.

\end{quotation}

\newpage
\pagestyle{plain}

\leftskip 3in
    \it  In the memory of Michael Polivanov, \\
    \it  the remarkable personality and scientist.

\leftskip 0pt
     \rm

\section{Introduction}

There are several reasons that make it desirable to improve mathematical tools
used in the nonlocal theory of highly singular quantum fields with an
exponential or faster high-energy behaviour. First of all string theory gives us
new indications [1-3] that the concept of space-time manifold is only
approximate and valid at length scales coarser than the Planck scale. Certainly,
one or another kind of nonlocality arises at any attempt to unify quantum
gravity with other fundamental interactions, however it is of great interest
that just the exponential growth is characteristic of the spectral densities
which occur in the K\"{a}llen-Lehmann representations for string propagators
with pointlike boundary conditions [4,5]. A second reason concerns the problem
of formulating causality which is crucial for any nonlocal theory. Recall
that the exponential bound on the off-mass-shell amplitudes has been found
by Meiman [6] just from the microcausality considerations. For faster
momentum-space growth, no definite criterion of macrocausality in terms of
observables has been obtained as yet. However a mathematical analog of local
commutativity has been found which ensures a number of important physical
consequences for arbitrary high-energy behaviour. These include the existence
of the unitary scattering matrix [7] and the polynomial boundedness of the
scattering amplitudes within the physical region of variables [8,9]. It is
surprising enough that the connection between spin and statistics and the
$TCP$-invariance derived previously [10-12] for quasilocal fields can also be
established [13-15] in the essentially nonlocalizable case when the holomorphy
domain of vacuum expectation values becomes empty and radically new proofs
are needed. Thirdly, the use of highly singular nonlocal form-factors [16]
turns out to be effective for a phonomenological description of strong
interactions [17-18].

The main purpose of the present paper is a more precise formulation of the
above-mentioned generalization of the local commutativity condition. This is
accomplished by adopting two ideas of the Sato-Martineau theory of
hyperfunctions which can briefly be named complexification and compactification
of space-time. This enables one to simplify significantly the mathematical
techniques used in the papers devoted to nonlocal quantum fields and develop
a convenient functional calculus effective beyond the theory of hyperfunctions [19].

The work is organized as follows. In Sec.2, we redefine in terms of complex
variables the Gelfand-Shilov test function spaces that used in the nonlocal
quantum field theory. In Secs.3 and 4 the key notions of a carrier-cone and
support cone are introduced for nonlocalizable distributions. The proof of
the existence of the support cones leads naturally to an asymptotic
commutativity condition formulated in Sec.5 as a fall-off property of the
field commutator in the spacelike directions. As an application,
generalization of the Paley-Wiener-Schwartz theorem to the nonlocalizable
case is presented in Sec.6. In Sec.7, a new derivation of the K\"{a}llen-Lehmann
representation for propagators of open and closed strings is proposed. It is
essential that this derivation is applicable to arbitrary initial and final
string configurations and shows the same character of causal and local
properties as that established in [4,5] under pointlike boundary conditions.
These considerations are somewhat heuristic and serve as a motivation and
background for the previous rigorous analysis. Sec.8 is devoted to concluding
remarks.

\section{Redefinition  of  the presheaf of Gelfand-Shilov \\ spaces}

In the framework of Wightman axiomatic approach [20], quantum fields are
treated as operator-valued distributions over the Schwartz space $S$. Wider
classes of quantum fields are obtained if $S$ is replaced by a dense subset
with a finer topology. For instance, the Jaffe strictly local fields [21],
whose vacuum expectation values are ultradistributions, are defined on the
Gelfand-Shilov spaces $S^\beta$, where $\beta >1$, with the
Schwartz space being roughly $S^\infty$ . The Meiman fields [6] are defined
on $S^1$ and correspond to hyperfunctions.  For still wider classes of
distributions the notion of support loses its sense and nonlocality comes into
play. The space $S^{1,b}$ is adequate to the exponential high-energy
behaviour characteristic of the quasilocal field theory investigated in
[8,10-12]. Although exactly this behaviour shows itself in string theories, we
wish to develop here a mathematical formalism covering so-called essentially
nonlocalizable fields defined on $S^\beta$ with $\beta <1$, which may be of
advantage in deriving the {\it TCP}-invariance and the connection between spin
and statistics by L\"{u}cke's method [13,14]. We recall [22] that $S^\beta$
consists of those complex valued infinitely differentiable functions that
satisfy the inequalities

$$
|x^{k} \partial ^q \varphi (x)|\/  \leq \/  C_{k}\/ b^{|q|}\/ q^{\beta q},
\eqno{(1)}
$$

\noindent
where $x\in {\bf R}^n$, the constants $C_k$ and $b$ depend on the function
$\varphi$; $k$ and $q$ are multi-indices, and the standard notation of the
theory of functions of several variables are used. Accordingly, $S^\beta$
can be made into a topological space by means of the projective limit $|k|
\rightarrow \infty$ and the inductive limit $b\rightarrow \infty$, proceeding
from the norms

$$
||\varphi||_{N,b}\/ =\sup_{|k|\leq N}{}_{x,q}\/
\frac{|x^{k}\partial^{q} \varphi(x)|}{b^{|q|}q^{\beta q}}
\eqno{(2)}
$$

\noindent
It perhaps should be noted that $\sup_{|k|\leq N}|x^k|$ can be replaced here
by $(1+|x|)^N$.  Besides $S^\beta$, it is advisable to use the spaces
$S^\beta (O)$, with $O$ an open set in  ${\bf R}^n$, which are defined in the
same manner but with $x\in O$ . Such a collection of spaces is customary
referred to as a presheaf. For $\beta <1$, every function belonging to
$S^\beta$  can be analytically continued into the whole of ${\bf C}^n$ as an
entire function whose order of growth less than or equal to $1/(1-\beta)$, see
[22]. We shall rewrite analogously the definition of $S^\beta (O)$ for
conelike domains $O$ which will play a leading role below. For this purpose, we
introduce one more functional space.

{\bf Theorem 1}. Let $U$ be an open connected cone in ${\bf R}^n$ and $d(x,U)$
be the distance from the point $x$ to $U$. For $\beta <1$, the space
$S^\beta (U)$ is isomorphic to the space $E^\beta (U)$ of entire functions
satisfying the inequalities

$$
|x^{k}  \varphi (x+i y)|\/ \leq \/  C_{k}\/  \exp \{\,  |by|^{\beta ^\prime
} +d(bx,U)^{\beta^\prime}\} \eqno{(3)}
$$

\noindent
where the designation $\beta^\prime = 1/(1-\beta)$  is used for brevity. (Note
that $\beta^\prime \geq 1$.)

{\sl Remark.\/} The choice of the norm in ${\bf R}^n$ is unessential here
because all these are equivalent and the inductive limit $b\rightarrow
\infty$ is implicit.  Furthermore $d(bx,U) = bd(x,U)$  since by a cone is meant
one with its vertex at the origin.

{\sl Proof.\/} Let us denote by $S^{\beta,b,N}(U)$ the Banach space with the
norm $\|\cdot\|_{U,N,b}$. Owing to the assumption $\beta <1$ the Taylor series
expansion of $\varphi \in S^{\beta,b,N}(U)$  is convergent for all $z\in {\bf
C}^n$ and since its center $\xi \in U$ can be taken arbitrarily, the
behaviour of the analytic continuation is estimated by

$$
\inf_{\xi \in U}\/ (1+|\xi|)^{-N} \sum_{q}b^{|q|} |(z-\xi )^{q}|q^{\beta q}/q!
\eqno{(4)}
$$

\noindent
We replace the sum by the supremum with a slightly greater $b$. An easy
evaluation shows that

$$
{\rm const}\cdot\exp\{|h^\prime z|^{\beta^\prime}\}\/  \leq \/
\sup_{q}|z^q|q^{\beta q}/q!\/ \leq \/ \exp\{|hz|^{\beta^\prime} \} \eqno{(5)}
$$

\noindent
where $|\cdot|$ is the Lebesque $l^{\beta^\prime}$-norm,
$h=e(e\beta^\prime)^{-1/\beta^\prime}$ and $h^\prime$ is arbitrarily near to
$h$.  The inequality  $|z|^{\beta^\prime}\leq
|2x|^{\beta^\prime}+|2y|^{\beta^\prime}$ enables the exponential be factorized.
Further, let $\xi_0$ be a point of $\overline{U}$ such that $d(x,U) =
|x-\xi_0|$ .  Then

$$
\inf_{\xi \in U}{}(1+|\xi|)^{-N} \exp \{\,  |x-\xi|^{\beta ^\prime } \}\/  \leq
\/  (1+|\xi_{0}|)^{-N} \exp \{ d(x,U)^{\beta ^\prime}\}
$$

\noindent
If $x\in U$, then $\xi_0 = x$ and $d(x,U) = 0$, whereas for the points $x$
whose distance to $U$ does not exceed some $\theta > 0$, we have $|\xi_0| \geq
(1-\theta)|x|$. For the remainder points, the inequality

$$
(1+|\xi_{0}|)^{-N} \leq \/   1\/ \leq C_{\varepsilon}\/  (1+|x|)^{-N} \exp \{
d(\varepsilon x,U)^{\beta ^\prime }\}
$$

\noindent
holds with arbitrarily small $\varepsilon >0$. Combining our estimates, we
see that $S^{\beta,b,N}(U)$ is continuously embedded into the Banach space
$E^{\beta,b^\prime,N}(U)$ of entire functions, with the norm

$$
\|\varphi\|_{U,N,b^\prime}^\prime=\/
\sup_{z}\, (1+|x|)^{N}\, |\varphi(z)|\exp\{-|b^\prime
y|^{\beta^\prime}-d(b^\prime x,U)^{\beta^\prime}\}, \eqno{(6)}
$$

\noindent
where $b^\prime>2hb$.

To prove the converse, let $\varphi \in E^{\beta,b,N}(U)$. Due to the
inequality

$$
1+|\xi|\leq (1+|x|)(1+|x-\xi|)
$$

\noindent
we have

$$
(1+|x|)^{-N}\, \exp \{d(b x,U)^{\beta ^\prime
}\}\/ \leq\,C_{N,b^\prime}\,\inf_{\xi \in U}(1+|x|)^{-N}\,\exp\{|b^\prime
(x-\xi)|^{\beta^\prime}\},
\eqno{(7)}
$$

\noindent
where $b^\prime$ is arbitrarily near
to b. Furthermore

$$
\exp\{\/ |x|^{\beta^\prime}+|y|^{\beta^\prime}
\}\/\leq\/\exp\{\/2\sum_{j}|z_{j}|^{\beta^\prime}\}
\eqno{(8)}
$$

\noindent
Let us fix $\xi \in U$ and apply the Cauchy formula for the polydisk

$$
D=\{z\in {\bf C}^n:\/ |z_j-\xi_j|\leq r_j\}.
$$

\noindent
Making use of (7), (8), we
obtain

\begin{align*}
(1+|\xi|)^{N}\,|\partial^{q}\varphi(\xi)|\,&=\,\frac{q!}{(2\pi)^n}
\left|\hspace{1mm} \int\limits_{\partial_{0}D} ^{}
\frac{(1+|\xi|)^{N}\varphi(z)\,dz}{(z-\xi)^{q+I}} \right| \/  \\
&\leq\/C_{N,b^\prime}\hspace{1mm} \|\varphi\|_{U,N,b}^\prime\hspace{1mm}
q!\hspace{1mm}r^{-q}\/\exp\{\/2\sum_{j}(b^\prime r_{j})^{\beta^\prime}\}
\end{align*}

\noindent
Now one can exploit the freedom in the choice of $r$ and take the lower bound,
which gives

$$
(1+|\xi|)^{N}\,|\partial^{q}\varphi(\xi)|\,\leq\,C_{N,b^\prime}\,
\|\varphi\|_{U,N,b}^\prime\hspace{1mm} q!\,(2b^\prime
e/h)^{|q|}\,q^{-q/\beta^\prime}.
$$

\noindent
Finally we use the Stirling formula and conclude that $E^{\beta,b,N}(U)$ is
continuously embedded into $S^{\beta,2b^\prime /h,N}(U)$. Taking the projective
and inductive limits completes the proof.

\section{Carrier cones}

Now we introduce test function spaces associated with closed cones in ${\bf
R}^n $.  Recall that by the projection of a cone is meant its intersection with
the unit sphere and that a cone $V$ is called a compact subcone of another cone
$U$ if $\overline{\mbox{pr}V} \subset \mbox{pr}U$ . Then the notation
$V\Subset U$ is commonly used. let $B$ be a ball in ${\bf R}^n$ centered about
the origin.  Given a closed cone $K$, we set

$$
S^\beta(K)\,=\,\lim_{\longrightarrow} S^\beta(U\cup B)
\eqno{(9)}$$

\noindent
where $U$ runs over open cones such that $K\Subset U$.

{\bf Definition 1}. Let $f\in (S^\beta)^\prime,\/ \beta<1$. A closed
cone $K\subset {\bf R}^n$ will be said to be a carrier cone of $f$ if this
linear form is continuous under the topology induced on $S^\beta$ by that of
$S^\beta(K)$.

Let us dwell on the motivation of this definition. For $\beta<1$, the space
$S^\beta(U\cup B)$ does not change with variations of the radius of the
ball $B$ whose role is only to provide connectedness. From the proof of Theorem
1, it is obvious that this space coincides with $E^\beta(U)$. It is
worthwhile to point out the particular case of the degenerate cone $K=\{0\}$.
Then $S^\beta(K)$ is identical to the space of entire functions satisfying
the bound (3) with $d(bx,U)$ replaced by $|bx|$ . The inductive limit (9) is
actually taken over special neighbourhoods of $K$ which can be regarded as
traces of the neighbourhoods of its closure in the radially compactified space
$\hat{{\bf R}}^n$. Using such a compactification, we follow to the paper [23]
where the local properties of the Fourier hyperfunctions was investigated. The
point is that for them there exist smallest carriers among compact sets in $\hat{{\bf R}}^n$,
while there are no minimal carriers in ${\bf R}^n$. We shall show that this
important fact holds true for distributions on $S^\beta$, $\beta<1$,
providing we restrict ourselves by those compact sets that are closures of
cones. It should be noted that there is a natural one-to-one correspondence
between the cones in ${\bf R}^n$ and the subsets of the compactifying sphere
and that the family of spaces $E^\beta(U)$ may be thought of as a presheaf
over this sphere which is really a sheaf in contrast to the initial
$S^\beta(O)$. Keeping this motivation at the back of our mind, we prefer to
say below about cones in ${\bf R}^n$ rather than about compact sets in $\hat{{\bf R}}^n$
 and
to use the customary designation $S^\beta(U)$ instead of $E^\beta(U)$ even
though we shall deal with the complex representation (3).

Definition 1 can be rewritten as a fall-off property of the distribution $f$
inside the complementary cone. To this end, we shall smooth $f$ by convolution
with test functions decreasing rapidly enough and we shall use one more
class of Gelfand-Shilov space, defining them in terms of complex variables
from the outset. Namely, the space $S_\alpha^\beta(U)$, where $\alpha >0,
\beta <1$, and $U$ is an open cone in ${\bf R}^n$, consists of all entire
functions with the property that

$$
|\varphi(z)|\,\leq\,C\,\exp\{\/|by|^{\beta^\prime}+d(bx,U)^{\beta^\prime}-
|x/a|^{1/\alpha}\}
\eqno{(10)}
$$
\noindent
Here the constants $a, b$, and $C$ depend on $\varphi$. Accordingly,
$S_\alpha^\beta (U)$ is equipped with the inductive limit topology by
$a,b\rightarrow \infty$. We recall [22], that this space is only nontrivial
provided $\alpha+\beta \geq 1$.

{\bf Theorem 2}. Let $f\in (S^\beta)^\prime$ and let $K$ be a carrier cone of
$f$. Suppose that $\varphi \in S_\alpha^\beta$. Then the convolution $(f\star
\varphi)(x)=(f(\xi),\/ \varphi(x-\xi))$ belongs to every space
$S_\alpha^\beta (C)$ with $C$ an open compact subcone of $\complement K$. Moreover
the mapping $S_\alpha^\beta \longrightarrow S_\alpha^\beta (C):\varphi
\longrightarrow f\star \varphi$ is continuous.

This theorem proven in [24] is valid for any $\beta \geq 0$, but here we
would like to present for $\beta \leq 1$ an alternative proof using the
isomorphism $S^\beta(K)=E^\beta(K)$. Let $\varphi \in S_{\alpha,a}^{\beta,b}$.
We proceed from the formula

$$
|(f\star\varphi)(z)|\,\leq\,\|f\|_{U,a^\prime,b^\prime}\,
\|\varphi(z-\zeta)\|_{U,a^\prime,b^\prime}
\eqno{(11)}
$$
\noindent
which holds for any $a^\prime, b^\prime$ and each cone
$U$ such that $K\Subset U$. By definition of the norms involved in (11), we
have

\begin{align*}
|\varphi(z-\zeta)\|_{U^\prime,a^\prime,b^\prime}\,&\leq\,\|\varphi\|_{a,b}
\sup_{\zeta=\xi+i \eta}\,\exp\{\/|b(y-\eta)|^{\beta^\prime}
-|(x-\xi)/a|^{1/\alpha} \\
&\qquad \qquad\qquad\qquad-|b^\prime y|^{\beta^\prime}-d(b^\prime
\xi,U)^{\beta^\prime}+|\xi/a^\prime|^{1/\alpha}\}.
 \end{align*}

\noindent
Let us show that if $a^\prime, b^\prime$ are sufficiently large and the cone $U$
is separated from $C$ by a nonzero angle distance, the right-hand side of this
inequality can be majorized by

$$
{\cal C}
\hspace{1mm}\|\varphi\|_{a,b}\,\exp\{\/|2by|^{\beta^\prime}+d(2b,C)^{\beta^\prime}
 -|\theta x/a|^{1/ \alpha}\}
\eqno{(12)}
$$
\noindent
In fact, in estimating the supremum over $\eta$ one can exploit convexity of
the function $|\cdot |^{\beta^\prime}$ and take $b^\prime \geq 2b$. In
considering the supremum over $\xi$, we assume at first that $x$ lies in the
cone $C_\theta=\{x:\/d(x,C)\leq \theta|x|\}$. If $\theta$ is small enough,
then

$$
|x-\xi|\,\geq\,2^\alpha\,\max(\/\theta |x|,\theta|\xi|\/) \hspace{5mm}
\mbox{ for all} \hspace{2mm}             x\in C_\theta,\,\xi\in U_\theta
$$
\noindent
and one can replace $|x-\xi|^{1/\alpha}$ by $|\theta x|^{1/\alpha}+|\theta
\xi|^{1/\alpha}$. For $\xi \bar{\in}U_\theta$, we have $|x/2|^{1/\alpha}\geq
|x-\xi|^{1/\alpha}+|\xi|^{1/\alpha}$ . Taking $a^\prime >a /\theta, b^\prime
> 2^\alpha/(a\theta)$, we reveal the decrease $\exp\{-|\theta
x/a|^{1/\alpha}\}$ within the cone $C_\theta$. If $x\bar{\in}C_\theta$, then
$|\varepsilon \xi|^{1/\alpha} \leq |x-\xi|^{1/\alpha}+|\varepsilon
x/(1-\varepsilon)|^{1/\alpha}$ as follows from the elementary inequality
$(\varepsilon u+(1-\varepsilon)v)^{1/\alpha} \leq u^{1/\alpha}+v^{1/\alpha}$
valid for positive $u, v$ due to the monotone behaviour of this function in
$\varepsilon \in [0,1]$. Choosing $\varepsilon$ and $a^\prime$ so that
$\varepsilon/(1-\varepsilon) < ab\theta,\theta^\prime>a/\varepsilon$, we obtain
a growth no faster than $\exp\{|b\theta x|^{1/\alpha}\}$ outside $C_\theta$.
It remains to notice that under the conditions $b>1/a, |x|>1/(b\theta)$ the
function equal to $-|\theta x/a|^{1/\alpha}$ on $C_\theta$ and $|b\theta
x|^{1/\alpha}$ elsewhere is bounded by the sum of the last two terms in the
exponent (12).

In particular, setting $\alpha=1-\beta$, we obtain

$$
|(f\star \varphi)(x)|\/\leq\/{\cal C}_{a,b}(C)\/
\|\varphi\|_{a,b}\,\exp\{-|\theta x/a|^{\beta^\prime}\} \hspace{5mm}  (x\in
C\Subset \complement K, \hspace{2mm}\varphi\in S_{\alpha,a}^{\beta,b}) \eqno{(13)}
$$

\noindent
This formula provides the basis for all the following study and makes evident
the meaning of the generalization of local commutativity presented in Sec.5.
In the next section, we shall show that this fall-off property is completely
equivalent to the fact that $f$ is carried by the cone $K$.

\section{Support cones}

{\bf Theorem 3}. Let $f\in (S^\beta)^\prime,0<\beta<1$, and let $K$ be a closed
cone in ${\bf R}^n$. Assume that for each cone $C\Subset \complement K$ there
exists a constant $\theta > 0$ such that the estimate (13) holds with any $a, b$.
Then $K$ is a carrier cone of $f$.

{\sl Proof.\/} We have to extend the linear form $f$ to every space.
$S^{\beta,b^\prime}(U)=\bigcap_N S^{\beta,b^\prime,N}(U)$ with $U$ an open cone
such that $K\Subset U$. The extension can be defined by the formula

$$
(\hat{f} ,\/\varphi)\/=\/\int (f(\zeta),\,\varphi_0(x-\zeta)\varphi(\zeta))\,
dx,
\eqno{(14)}
$$

\noindent
where $\varphi_0\in S_{1-\beta,a_0}^{\beta,b_0},\/\/ \int \varphi_0(x)\/ dx=1$
and $a_0$, $b_0$ will be chosen below. Let us denote the integrand by $J(x)$,
fix a cone $V$ so that $K\Subset V \Subset U$, and consider at first the
behaviour of $J(x)$ inside $V$. For each $b$, $N^\prime$ and some $N$ depending
on $b$,

$$
\begin{array}{c}
|J(x)|\/\leq\/\|f\|_{N,b}\hspace{1mm}\|\varphi_0(x-\zeta)
\varphi(\zeta)\|_{N,b}\,\leq\\    [9pt]
\leq\,\|f\|_{N,b}\hspace{1mm}\|\varphi_0\|_{a_0,b_0}\,
\|\varphi\|_{U,N^\prime,b^\prime}\,
\sup_{\xi,\eta}(1+|\xi|)^{N-N^\prime}\times  \\  [9pt]
\times \exp\{\/|b_0 \eta|^{\beta^\prime}
-|(x-\xi)/a_0|^{\beta^\prime}+|b^\prime \eta|^{\beta^\prime} + d(b^\prime
\xi,U)^{\beta^\prime} -|b\eta|^{\beta^\prime}\} \end{array} \eqno{(15)}
$$

\noindent
If $\xi \bar{\in} U$, we have $|x-\xi|^{\beta^\prime}\geq |\theta^\prime
x|^{\beta^\prime}+|\theta^\prime \xi|^{\beta^\prime}$ since the cone
 $\complement U$
is separated from $V$ by a nonzero angle distance. For $\xi \in U$, we use the
inequality $1+|x|\leq (1+|x-\xi|)(1+|\xi|)$. Choosing
$a_0<\theta^\prime/b^\prime$ and $b>2(b_0+b^\prime)$, we see that $J(x)$
decreases within $V$ faster than any inverse power of $|x|$. In estimating
this function outside $V$ we regard it as $f\star\chi_x$, where
$\chi_x(\zeta)=\varphi_0(\zeta)\varphi(x-\zeta)$, denote $\complement V$ by $C$ and
apply (13). This time we deal with the $S_{1-\beta,a}^{\beta,b}$-norm of
$\chi_x$ which involves

$$
\sup_{\xi}(1+|x-\xi|)^{-N^\prime}\,\exp\{-|\xi/a_0|^{\beta^\prime} +
d(b^\prime(x-\xi),U)^{\beta^\prime} + |\xi/a|^{\beta^\prime}\}.
\eqno{(16)}
$$

\noindent
(We do not write out the supremum over $\eta$, assuming $b$ subject to the
above condition.) Using the inequality $d(\cdot ,U)\leq |\cdot|$ and choosing
$a_0<1/(2b^\prime),a>2a_0$, we find that $\|\chi_x\|_{a,b}$ increases no faster
than $\exp\{|b^\prime x|^{\beta^\prime}\}$. Hence, if $a<\theta /b$, then
$J(x)$ decreases exponentially everywhere outside the cone $V$. Thus, the
integral (14) is convergent under the condition
$a_0<\min(\theta^\prime,\theta/2)/\beta^\prime$ and determines a linear form
on $S^{\beta,b^\prime}(U)$ whose continuity is ensured by the presence of the
factor $\|\varphi\|_{U,N^\prime,b^\prime}^\prime$ in our estimates. This
distribution coincides with the initial one, when restricted to $S^\beta$.
In fact, let $\varphi \in S^\beta$. Then the term $d(b^\prime\xi,U)$
is absent from (15) and this estimate shows that not only $J(x)$ but also the
function $(g(\zeta),\/ \varphi_0(x-\zeta)\varphi(\zeta))$ is integrable for
each $g\in(S^\beta)^\prime$. Therefore the sequence of Riemann sums
corresponding to the integral $\int \varphi_0(x-\zeta)\varphi(\zeta)\/ dx$ is
weakly fundamental in $S^\beta$ and, this space being Montel, converges in
its topology to some element which is of necessity $\varphi$. Now we recall
[22] that the space $S_{1-\beta,a}^{\beta,b}$ is only nontrivial provided
$ab\geq \gamma$ where $\gamma$ depends on $\beta$. Hence different
$\varphi_0$ occur in the formula (14) for different $b^\prime$ and we have to
make sure of consistency of all the extensions.

{\bf Lemma}. Let $U$ be an open cone in ${\bf R}^n$ and $0<\beta<1$ as
before. The space $S_{1-\beta}^\beta$ is dense in $S^\beta(U)$.

It perhaps should be recalled that we always add a ball $B$ to the cone to
ensure connectedness. Let $\varphi\in S^{\beta,b^\prime}(U)$ and let
$\varphi_0\in S_{1-\beta,a_0}^{\beta,b_0}$, where $a_0=1/(2b^\prime)$ and
$b_0=\gamma/a_0$. The estimate (16) with $\xi$ replaced by $x-\xi$ shows
that, for each $x$, the function $\varphi_0(x-\xi)\varphi(\xi)$ belongs
to $S_{1-\beta,a}^{\beta,b}$ with $a>1/b^\prime$ and $b>2(b_0+b^\prime)$.
The above-mentioned sequence of Riemann sums lies in this space too. Moreover it
is weakly fundamental in $S^{\beta,b}(U)$ as can be seen from a norm estimate
similar to (15). Hence it converges to $\varphi$ in the topology of every
space $S^{\beta,b_1}(U)$ with $b_1>b$ since their intersection is a Montel
space. This completes the proof of Theorem 3.

{\bf Corollary}. For each distribution $f$ defined on $S^\beta,\/ 0<\beta<1$,
there exist a smallest carrier cone which can be called the support cone of $f$.

{\sl Proof.\/} Let us denote by $K$ the intersection of all the carrier cones
and consider an open cone $C\Subset \complement K$. The complements of the carrier
cones form an open covering of $C$ from which one can choose a finite
subcovering. Slightly shrinking the cones involved in the latter, we obtain
covering cones where (13) is satisfied. Therefore this estimate holds everywhere
in $C$ with some nonzero $\theta$ and so $K$ is a carrier too.

\section{Asymptotic commutativity}

Now we are in a position to formulate the following.

{\bf Definition 2}. Let $\{A^{(\kappa)}\}$ be a finite or infinite set of
nonlocal quantum fields defined on the test function space $S^\beta({\bf
R}^n),\hspace{2mm} 0<\beta <1$, and transforming according to finite-dimensional
representations of the proper orthochronous Lorentz group . We say Lorentz
components $A_j^{(\kappa)}$ and $A_{j^\prime}^{(\kappa^\prime)}$ are
asymptotically (anti)commute if for each vectors $\Phi,\Psi$ from their common
domain in the Hilbert space, the distribution

$$f=\langle
\Phi,\/[A_j^{(\kappa)}(x),
 A_{j^\prime}^{(\kappa^\prime)}(x^\prime)]_{\stackrel{-}{(+)}} \Psi \rangle $$

\noindent
is carried by the cone $\overline {{\sf
V}}=\{(x,x^\prime):\/\/(x-x^\prime)^2\geq 0\}$ or, equivalently, fulfills the
condition (13) outside this cone. The principle of asymptotic commutativity
means that every two field components asymptotically commute or anticommute. We
assume that the type of this relation depends only on the type of the fields,
{\it i.e.\/}, on $\kappa,\kappa^\prime$.

It should be mentioned that the term "asymptotic commutativity" has been
introduced in Ref.[25] but with some other meaning. Namely, L\"{u}cke called
nonlocal tempered fields $A, A^\prime\hspace{2mm} \varphi$-asymptotically
commuting, where $\varphi \in S({\bf R}^n)$ and decreases rapidly enough, if
the function

$$
F(x)=\left\|\int d\xi\/ d\xi^\prime\/ [A(\xi),A^\prime(\xi^\prime)]\/
\varphi(x-\xi)\varphi(\xi)\/\Phi\right \|
$$

\noindent
has finite the
norm $\|F\|_{\lambda,\varepsilon}=\sup\{|F(x)|\exp(\varepsilon|x|):\/ |{\bf
x}|>\lambda x^0\}$  for all $\lambda,\varepsilon>1$. Generalizations of the
principle of local commutativity to fields breaking the Jaffe strict
localizability condition [21] were considered by several authors [6-18, 26-28].
A formulation exploiting the presheaves of Gelfand-Shilov spaces and the
requirement of continuity under an induced topology has been proposed in [26],
but there the main attention was paid to the conditions under which this
generalization keeps sense of microcausality. In [9], this approach was applied
to nonlocal fields and the triviality of $S_\alpha^\beta$ for $\alpha+\beta<1$
was interpreted is such a way that the functions belonging to $S^\beta$ enable
one to test the causal properties to an accuracy of
$\exp\{-|x-x^\prime|^{1/(1-\beta)}\}$. Theorems 2 and 3 confirm this view.
Closely related definitions have been proposed independently in [7,27].
B\"{u}mmerstede and L\"{u}cke [7] suggested to replace the local commutativity
by an axiom of "essential locality" which implies continuity of the field
commutator under the topology of $\displaystyle\lim_{\longrightarrow} S^\beta
\cap S^{\beta,b}({\sf V})\/ \/ (b\rightarrow \infty)$.  Constantinescu and
Taylor [27] introduced a classification of extensions of the field commutator
outside the light cone.  Namely, for a field $A$ on $S^{\beta_0}$, they defined
the order of the extension to be the greatest $\beta^\prime=1/(1-\beta)$ such
that all the distributions $\langle \Phi,\/ [A(x),A(x^\prime)]\Psi \rangle$ can
be extended to $S^\beta({\sf V})$. A comparison of these two formulations can be
found in [28]. The papers [7,9,26-28] proceeded from the definition (1), (2)
while Efimov [17] took advantage of the redefinition of $S^\beta$ in terms of
complex variables. The spaces (9) and the idea of compactification were used
neither in our or other above-mentioned papers. From the technical standpoint
the compactification procedure means taking an additional inductive limit over
conelike neighborhoods of ${\overline {\sf V}}$. It is worthwhile to point out
that the continuity under this topology is a weaker requirement than that of
continuity under the topology of $S^\beta({\sf V})$. It is the compactification
that enables one to derive Theorem 3, to establish the existence of support
cones and to prove the equivalence of the two versions of asymptotic
commutativity formulated as a support property and as a fall-off property of the
field commutator.

\section{Generalization of the Paley-Wiener-Schwartz \\ theorem}

It will be recalled that this theorem [29] relates the support properties of
tempered distributions to the growth properties of their Laplace transforms.
In [30], it has been extended to the ultradistributions defined on
$S_\alpha^\beta,\/ \/\beta >1$.  A generalization to the nonlocalizable
case $\beta < 1$ was also formulated there, but without a proof\footnote{The
case of Fourier hyperfunctions $\alpha=\beta=1$ had been considered by Kawai
[23].}. The results of Sec.4 enables us to present such a proof here. It is
worthwhile to note that this generalization, interesting by itself, is needed to
derive the connection between spin and statistics for nonlocalizable fields, see
[13,15].

{\bf Theorem 4}. Let $K$ be a closed convex cone in ${\bf R}^n$ which does not
contain a straight line, and let $V$ be the interior of its dual cone
$K^*=\{q:\/\/ qx\geq 0,x\in K\}$.  Suppose that $K$ is a carrier of $f\in
(S^\beta)^\prime,\/\/ 0<\beta<1$. Then for the distribution $f$ there exists a
Laplace transform $g(s) (s=p+i q)$ holomorphic in the tubular cone
$T^V={\bf R}^n +i V$ and satisfying the inequality

$$
|g(s)|\,\leq\,C_\varepsilon(V^\prime))\,|q|^{-N}\,\exp\{\/|\varepsilon
s|^{1/\beta}\}\hspace{5mm}  (q\in V^\prime)  \eqno{(17)}
$$

\noindent
for any $\varepsilon>0\/\/ ,V^\prime\Subset V$ and some $N$ depending on
$\varepsilon,V^\prime$. Conversely, if $g$ is a function holomorphic in a tube
$T^V$, with $V$ an open connected cone, and satisfying the growth condition
(17), then it serves as the Laplace transform of a distribution
$f\in (S^\beta)^\prime$ carried by the cone $V^*$.

{\sl Proof.\/} We shall show that $e^{i sz} \in S^\beta(K)$ for any $q\in
V=\hbox{int} K^*$ and then define the Laplace transformation by the formula

$$
g(s)\/=\/(f(z),\/e^{i sz})
\eqno{(18)}
$$

\noindent
After that the estimate (17) can be derived from the inequality

$$
|g(s)|\,\leq\,\|f\|_{U,N,b}\,\|e^{i sz}\|_{U,N,b}
$$

\noindent
where $K\Subset U\/\/ , b$ is arbitrarily large and $N$ depends on $U, b$. We
have

$$
\|e^{i sz}\|_{U,N,b}\,=\,\sup_{z}(1+|x|)^N
\,\exp\{-qx-py-|by|^{\beta^\prime}-d(bx,U)^{\beta^\prime}\}
\eqno{(19 )}
$$

\noindent
The inclusion $V^\prime\Subset V$ implies $K\Subset \hbox{int}V^{\prime *}$
. We choose the cone $U$ and another  open auxiliary cone $U^\prime$ so that
$K\Subset U\Subset U^\prime\Subset \hbox{int}V^{\prime *}$. In evaluating the
supremum over $y$ we use the inequality $-py\leq |p||y|$ and obtain the
factor $\exp\{|\varepsilon p|^{1/\beta}\}$ where $\varepsilon$ is arbitrarily
small.  Next we estimate the supremum over $x$ inside and outside the cone
$U^\prime$.  There exists $\theta > 0$ such that $-qx\leq -\theta|q||x|$
for all
$q\in V^\prime, \/ x\in U^\prime$. Indeed, the function
$i_{V^\prime}(x)=\sup\{-qx:\/ q\in \hbox{pr}V^\prime\}$ called the indicatrix
of $V^\prime$ is continuous negative in $\hbox{int}V^{\prime *}$ and takes
the maximal value $-\theta$ on the compact set $\hbox{pr}\overline{U}^\prime$.
This gives the factor $|q|^{-N}$, while for $x\in U^\prime$ we have
$d(x,U)>\delta|x|$, $\delta>0$ and so obtain the factor $\exp\{|\varepsilon
q|^{1/\beta}\}$.  Further, it can be easily verified that the difference
quotients corresponding to the partial derivatives $\partial e^{i
sz}/\partial s_j$ converge to them in the topology of $S^\beta (U)$ and so
all the derivatives $\partial g/\partial s_j$ exist. By the Hartogs theorem,
this means that $g$ is holomorphic in $T^V$.

Conversely, given a function $g$ analytic in $T^V$ and satisfying (17), we
define $f\in (S_{1-\beta}^\beta)^\prime$ by the formula

$$
(f,\,\varphi)\,=\,\int g(p+i q)\,\psi(p+i q)\,dp \hspace{5mm}  (q\in
V)
\eqno{(20)}
$$

\noindent
where $\psi(p)=(2\pi)^{-n}\int \varphi(x)e^{-i px}\/ dx$. Taking into
account that the Fourier operator is an isomorphism from $S_{1-\beta}^\beta$
onto $S_\beta^{1-\beta}$ (and maps $S_{1-\beta,a}^{\beta,b}$ into
$S_{\beta,b^\prime}^{1-\beta,a^\prime}$ with $a^\prime,\/  b^\prime$ arbitrarily
near to $a,\/  b$) and using (10), (17), we see that the right-hand side of (20)
is meaningful and independent of $q$. Then we consider the convolution

$$
(f\star \varphi)(x)\,=\,\int g(s)\,\psi(-s)\,e^{-i sx}\,ds\hspace{5mm}
(\hbox{Im} s\,=\,q\in V)
$$

\noindent
and exploit the latitude in the choice of $q$, which gives

$$
|(f\star\varphi)(x)|\,\leq\,C_{\varepsilon,b}(V^\prime)\,
\|\psi\|_{a,b}\,\inf_{q\in V^\prime} |q|^{-N}\,
\exp\{\/|\varepsilon q|^{1/\beta}+|aq|^{1/\beta}+qx\}.
$$

\noindent
Evaluating the infimum under the additional condition $|q|\leq 1$, we obtain
the following polynomial bound

$$
|(f\star\varphi)(x)|\,\leq\,C_{a,b}\,\|\psi\|_{a,b}\,(1+|x|)^N
\eqno{(21)}
$$

\noindent
Here $N$ depends on $b$ since $\varepsilon$ must be sufficiently small
compared to the number $1/b$ which characterizes the rate of decrease of
$\psi$. The continuity of the Fourier operator allows $\|\psi\|$ to be
replaced by $\|\varphi\|$ in the inequality (21). Applying the trick (14)
again, we make sure that this restriction on the growth enables the
distribution $f$ to be extended to the space $S^\beta$. Next we fix a cone
$C\Subset \complement V^*$ and estimate the infimum over $q\in V^\prime,\/ |q|\geq
1$. The cone $V^\prime$ can be taken so that $C\Subset \complement V^{^\prime *}$.
Then there exists $\theta^\prime > 0$ such that $qx\leq -\theta^\prime|q||x|$
for all $q\in V^\prime,\/ x\in C$.  Namely, $\theta^\prime$ is the minimal value
of the indicatrix $i_{V^\prime}(x)$ on the compact set
$\hbox{pr}\overline{C}$.  Finding this extremum, we arrive at the estimate
 (13) with $\theta$ arbitrarily near to $\theta^\prime
\beta^{\beta}(1-\beta)^{1-\beta}$ . According to Theorem 3 and Lemma, $f$ has a
unique extension to a distribution $\hat{f}$ on the space $S^\beta(V^*)$. The
exponential $e^{i sz}$ belongs to this space for any $\hbox{Im}s\in
\hbox{ch} V$, where ch signifies the convex hull, since
$V^{**}=\overline{\hbox{ch}V}$. Approximating the exponential by $\varphi_\nu
\in S_{1-\beta}^\beta$ and passing to the limit $\nu\rightarrow \infty$ in (20),
we see that the initial function $g$ and $\hat{f}$ are connected by the formula
(18). This completes the proof.

{\bf Corollary}. If $g$ is holomorphic in $T^V$ and fulfills the condition
(17), then it can be analytically continued into the tubular cone
$T^{\hbox{ch}V}$ where an analogous estimate holds.

Note that if $\hbox{ch}V={\bf R}^n$, then $V^*=\{0\}$ and the function $g$ is
entire analytic. In this case the term $d(bx,U)$ in (19) should be replaced by
$|bx|$ and an evaluation of extremum shows that (17) holds in the whole of
${\bf C}^n$ without the factor $|q|^{-N}$ of course.

\section{Causality and K\"{a}llen-Lehmann-like representations \\
for propagators of open and closed strings}

As shown in [12], the K\"{a}llen-Lehmann representation for particle
propagators in Min\-kow\-ski space-time holds true up to the exponential growth
of the spectral density: $\rho(M^2) \leq C\/ e^{lM}$. By the Meiman criterion,
the theories with such a growth are nonlocalizable, with $S^{1,1/l}$ being the
adequate test function space and $l$ playing the role of an elementary
length. Analogous representations for string propagators with the pointlike
boundary conditions were considered in [4,5].As would be expected [32], the
growth of their associated spectral density turns out to be exponential with the
fundamental length $l_{{\rm Pl}} \approx \sqrt{\alpha^\prime}$. In
investigating the string propagators, one usually employs the operator formalism
or the Polyakov method [32]. In this section, we present another approach
closely related to the secondary quantization.It is applicable to arbitrary
initial and final string configurations and shows that the character of causal
and singular properties of the string propagators is independent of the boundary
conditions. We shall set forth the main steps of the proof for the case of open
bosonic string and outline final results for closed ones. We shall follow the
notation of [34] and use in intermediate laying out the system of units
for which $\sqrt{2\alpha^\prime}=1$.

Let us proceed from the following BRST-invariant expression for the Hamiltonian
of the open bosonic string in $D$-dimensional space-time

$$
{\bf H}={\bf L}_0-1=\frac{1}{2}({\bf p}^2+{\bf M}^2),\quad {\bf
M}^2=2\sum_{n=1}^\infty n\left({\bf a}_{-n}^\mu {\bf a}_{n\mu}+{\bf c}_{-n}{\bf
b}_n+{\bf b}_{-n}{\bf c}_n\right) -2
\eqno{(22)}
$$

\noindent
(In fact we take $D$ to be equal to 26.) Here $p_\mu=-i \partial/\partial
x_\mu$ is the momentum operator canonically conjugate to the string center mass
position and playing a significant part in our approach. The
operators $a_{\pm n}^\mu$ are Fourier modes of the string coordinates
$X^\mu(\tau,\sigma)$ and $c_{\pm n},\/b_{\pm n}$ are ghost and antighst modes
satisfying the usual anticommutation relations. Keeping in mind the
zeta-function regularization, one might replace the normal ordering constant
$-2$
by $ (D-2)\sum_{n=1}^\infty n$, restrict the summation in (22) by a large $N$,
and then treat $H$ as a limit of Hamiltonians describing dynamic systems with
finite number of degrees of freedom. This would give a more accuracy to the
following study and put off the appearance of the tachyon but we will not go
into such details for brevity. We use the Berezin holomorphic representation
which is best suited to the Euclidean continual integretion and normalize the
corresponding measure to unity, {\it i.e.\/},

$$
\int d\omega=\prod_{n}\int \frac{da_{-n}\/da_n}{2\pi
i}e^{-a_{-n}a_n}\/\int dc_{-n}\/db_n \/db_{-n}\/dc_n \/e^{-c_{-n}b_n
-b_{-n}c_n}\/=\/1.
$$

\noindent
The index $\mu$ will henceforth be omitted and the notation $A$ will be used for
the collection of variables $\{a_1,b_1,c_1,\ldots,a_n,b_n,c_n,\ldots \}$. Let us
denote by $\delta(x,A^*;x^\prime,A^\prime)$ the kernel of unity operator defined
by

$$
\delta(x,A^*;\/x^\prime,A^\prime)=\delta(x-x^\prime)\sum_{k}\frac{(A^k)^*A^{\prime k}}{k!}
\eqno{(23)}
$$

\noindent
where $k$ runs through the multi-indices which have only finite number of
nonzero components $k_n$, each being $(D+2)$-tuple of nonnegative
integers. Of course, the indices corresponding to the Grassmann variables $b_n$
and $c_n$ make a contribution to the sum only being equal to 0 or 1. The
operation * consisting in the conjugation $a_n\rightarrow a_{-n}, b_n\rightarrow
c_{-n}, c_n\rightarrow b_{-n}$ implies the inversion of the order in the
products. The Green function for the operator $\bf{H}$ is defined by

$$
{\bf H}G(x,A^*;\/x^\prime,A^\prime)=\delta(x,A^*;\/x^\prime,A^\prime)
$$

\noindent
where $H$ acts on the former pair of variables, while the latter is regarded as
parameters. One can readily verify that

$$
G(x,A^*;\/x^\prime,A^\prime)=\int \frac{dp}{(2\pi)^{\rm
D}}\/2\sum_{k}\frac{e^{\imath
p(x-x^\prime)}}{p^2+2[k]-2}\/\/\/\frac{(A^k)^*A^{\prime k}}{k!}
\eqno{(24)}
$$

\noindent
where [k] stands for $\sum_{n}n|k_n|$. Using the summation over the spectrum of
the mass operator, one may reduce this representation to the K\"{a}llen-Lehmann
form

$$
G(x,A^*;\/x^\prime,A^\prime)=\sum_{M^2} \/\int
\frac{dp}{(2\pi)^{\rm D}}\/\/2\frac{\rho (M^2;\/A^*,A^\prime)}{p^2 +M^2}
$$

\noindent
where the spectral density matrix is determined by

$$
\rho (M^2;\/A^*,A^\prime)=\int\limits_{-1/2}^{1/2}
d\varphi\/e^{\pi i\varphi(M^2+2)}\/\sum_{k}e^{-2\pi i\varphi
[k]}\prod_{n}\frac{(A_n^{k_n})^*A_n^{\prime k_n}}{k_{n}!}=
\eqno{(25)}
$$
$$
=\int\limits_{-1/2}^{1/2}
d\varphi\/e^{\pi i\varphi(M^2+2)}\prod_{n}\sum_{k}e^{-2\pi i\varphi
[k]}\/\/\/\frac{(A_n^{k_n})^*A_n^{\prime k_n}}{k_{n}!}
\eqno{(26)}
$$

\noindent
Note that the $\varphi$-integration yields the Kronecker symbol since
eigenvalues of ${\bf M}^2$ are even integers. From (26), one can immediately
obtain an explicit expression for the spectral density

$$
\rho(M^2)=\hbox{Sp}\/\rho(M^2;\/A^*,A^\prime)=\int d\omega\/\/\rho(M^2;\/A^*,A)
$$

\noindent
in the case of pointlike boundary conditions. Namely,

$$
\rho(M^2)=\int\limits_{-1/2}^{1/2}
d\varphi\/e^{\pi i\varphi(M^2+2)}\/\prod_{n} \left(1-e^{-2\pi i n
\varphi}\right)^{-24}
\eqno{(27)}
$$

\noindent
It perhaps is worthwhile to note that the integration over $a_{-n},a_n$
 gives the power $-26$ which is reduced to $-24$ by virtue of the
integration over the Grassmann variables. Let us denote by $f(e^{-2\pi i
\varphi}) $ the product that occurs in (27). Substituting the power series
expansion $f(w)=\sum_{m} d_m w^m$ and performing the $\varphi$-integration, we
obtain the equality $\rho(M^2)=d_{M^2 /2+1}$. The asymptotic behavour of the
coefficients $d_m$ is well known [34, Sec.2.3.5] and we arrive at the formula

$$
\rho(M^2)\approx M^{-25/2}\exp(M/M_0)
\eqno{(28)}
$$

\noindent
where $M_0=1/(4\pi \sqrt{\alpha^\prime})$. It should be pointed out that
the representation (25) enables one to write the function $\rho$ in another
form

$$
\rho(M^2)=\sum_{k}\delta_{M^2 +2}^{2[k]}
\eqno{(29)}
$$

\noindent
expressing it as the degree of degeneracy of the mass operator eigenvalues. From
this viewpoint the problem of deriving the formula (28) looks pure
combinatorial. Thus in the case of pointlike initial and final string
configurations, the growth of the spectral density is indeed exponential. For
more general boundary conditions, one can make use of the following simple
facts. Let $R$ and $\rho$ be operators in a Hilbert space. Assume that $R$ is
bounded, {\it i.e.,\/} $\|R\|=\sup_{\varphi\neq 0}\|R\varphi\|/\|\varphi\|<\infty$, and
$\rho$ is compact. Then the positive self-adjoint operator$(\rho^*\rho)^{1/2}$
is also compact.
Let $\lambda_m$ be its eigenvalues. The nuclear norm is defined by $\|  \rho
\|_1 = \sum_m \lambda_m$ and a compact operator is called nuclear if this norm
is finite. The inequalities $| {\rm Sp}\rho | \le \| \rho \|_1$ and
$\|R\rho\|_1 \le \| R \| \|\rho \|_1$ are valid. It is easy to see that in our
case the nuclear norm of $\rho (M^2;A^*,A')$ is precisely the sum (29). It
follows that the spectral density  ${\rm Sp} R\rho$ behaves exponentially as
well for any bounded $R$ in the K\"{a}llen-Lehmann-like representation.

        The next step in studying the string propagator causal properties is
passing to the Minkowski space-time by means of analytic continuation in $p_0$.
Strictly speaking, this should be performed before the limit $N \rightarrow
\infty$ which breaks down microcausality due to the appearance of the tachyon
eigenvalue $-1/\alpha'$ in the spectrum of ${\bf M}^2$. On doing so, one can see
that the K\"{a}llen-Lehmann representation for string propagators make no sense
inside the hyperboloid $(x - x')^2 \approx \alpha'$ where singularities  present
themselves independently of the boundary conditions. It is natural to assume that
this conclusion is unaffected by taking into account the string interaction. For
the closed bosonic string, analogous formulae can be derived. This time,
however, we deal with right and left modes, {\it i.e.,\/} double of the variables and
the $\delta$-symbol must satisfy the additional condition $({\bf L}_0 -
\tilde{{\bf L}}_0)\delta = 0$ because of which formula (24) is replaced by

$$
G(x,A^*,\tilde{A}^*; \/x^\prime, A^\prime, \tilde{A}^\prime) = \int
\frac{dp}{(2\pi)^{\rm D}}\/\/2\sum_{k,\tilde{k}}\delta_{[k]}^{[\tilde{k}]}
\/\/\frac{e^{i p(x-x^\prime)}}{p^2+4[k]-4}\frac{(A^{k}
\tilde{A}^{\tilde{k}})^{*} A^{\prime
k}\tilde{A}^{\prime\tilde{k}}}{k!\tilde{k}!}
\eqno{(30)}
 $$

\noindent
and the spectral density matrix acquires the form

$$
\rho_{\hbox{cl}}(A^*,\tilde{A}^*;\/A^\prime,\tilde{A}^\prime)=
\int\limits_{-1/2}^{1/2}
d\varphi\/e^{\pi i\varphi(M^2+4)}\/\sum_{k,\tilde{k}}
\delta_{[k]}^{[\tilde{k}]}\/\/\/e^{-4\pi i\varphi [k]}
\/\/\frac{(A^{k}\tilde{A}^{\tilde{k}})^{*} A^{\prime k}\tilde{A}^{\prime
\tilde{k}}}{k!\tilde{k}!}
$$

\noindent
In this case, too, the exponential asymptotic behaviour takes place
independently of the initial and final string configurations and so the causal
and singular properties of the open  and closed string propagators are
practically identical. We would like to point out that in the one loop
approximation  the representation (29), making allowance for the constraint
${\bf L}_0 = \tilde{{\bf L}}_0$, leads immediately to the formally modular
invariant expression [35] for the cosmological constant. The approach presented
here can be extended to the open and closed fermionic Neveu-Schwarz-Ramond
strings and the final results are qualitatively the same but their derivation
will be discussed in some detail elsewhere.

\section{Concluding remarks}

        In this paper, we have no attempt to prove the existence of support
cones for those distributions which are defined on the space $S^0$ whose Fourier
transform is the Schwartz space ${\cal D}$. Such a proof would be extremely
desirable because this space consisting of functions with compact momentum-space
support is universal for nonlocal quantum fields and one imposes no restrictions
on the high-energy behaviour when incorporates it into the Wightman scheme. The
problem is unfortunately complicated by triviality of the space $S^0_1$ which
makes inapplicable the foregoing elementary methods aimed at clarity in
presenting the concept of asymptotic commutativity. However the modern theory
of functions of several complex variables enables one to solve it as we hope
to show before long. The most interesting applications of the mathematical
techniques developed here are improvements in L\"{u}cke's derivations [13,~14]
of the connection between spin and statistics and the {\it TCP}-invariance for
essentially nonlocal quantum fields. This topic is covered in considerable
detail in Ref. [36]. Of course, there are many unanswered questions concerning
the existence of a minimum spatial resolution in the string theory whereas
string interactions look local and this fact is commonly believed to be enough
to preserve causality. One of the principal features of the superstring theory
is the remarkable role of the modular invariance in the banishment of the
ultraviolet divergences. For this reason the main omission of the above
consideration seems to be lack of any examination how this invariance could
affect the causal and singular properties of closed string propagators. However
this is probably somewhat premature because so far there has been no consistent
formulation of the closed string field theory developed in such a manner that
the integration over the modular parameters is naturally restricted to the
fundamental region of the modular group. Another question of great importance is
the relation between the superstring theory and the effectively nonlocal quantum
theory of fields with infinitely many components and arbitrarily high spins.
These two approaches might turn out to be complementary standpoints on the
evolution of Universe, describing its different phases, and this was one of the
main inspirations of our investigation.


\end{document}